\documentclass{amsart}
\usepackage{amsmath}

\usepackage{mathtools}
\usepackage{amssymb}

  \usepackage{paralist}
  \usepackage{graphics} 
  \usepackage{epsfig} 
\usepackage{graphicx}  \usepackage{epstopdf}
 \usepackage[colorlinks=true]{hyperref}
\hypersetup{urlcolor=blue, citecolor=red}

\newtheorem{theorem}{Theorem}[section]

\newtheorem{lemma}[theorem]{Lemma}
\newtheorem{proposition}{Proposition}

\theoremstyle{definition}
\newtheorem{definition}[theorem]{Definition}
\newtheorem{remark}{Remark}

\newtheorem{assu}[theorem]{Assumption}
\newtheorem{ex}[theorem]{Example}

\usepackage{tikz}
\usetikzlibrary{arrows,shapes,positioning}
\usetikzlibrary{decorations.markings}
\tikzstyle arrowstyle=[scale=1]
\tikzstyle directed=[postaction={decorate,decoration={markings,
    mark=at position .65 with {\arrow[arrowstyle]{stealth}}}}]
\tikzstyle reverse directed=[postaction={decorate,decoration={markings,
    mark=at position .65 with {\arrowreversed[arrowstyle]{stealth};}}}]

\renewcommand{\v}{\mathrm{v}}
\renewcommand{\u}{\mathrm{u}}
\newcommand{\w}{\mathrm{w}}
\newcommand{\V}{\mathrm{V}}

\newcommand{\R}{\mathbb{R}}

\newcommand{\e}{\mathrm{e}}
\newcommand{\E}{\mathrm{E}}
\newcommand{\Co}{\text{Conv}}

\title[Convex and quasiconvex functions in metric graphs] 
      {Convex and quasiconvex functions in metric graphs}

\author[L. M. Del Pezzo, N. Frevenza and J. D. Rossi]{}


 \email{ldpezzo@dm.uba.ar}
 \email{nfrevenza@fing.edu.uy}
 \email{jrossi@dm.uba.ar}

\thanks{
L.D.P. and J.D.R. partially supported by 
CONICET grant PIP GI No 11220150100036CO
(Argentina), PICT-2018-03183 (Argentina) and UBACyT grant 20020160100155BA (Argentina). 
J.D.R. is also supported by MINECO MTM2015-70227-P (Spain).
}

\begin{document}
\maketitle

\centerline{\scshape Leandro M. Del Pezzo}
\medskip
{\footnotesize
 \centerline{CONICET and Departamento  de Matem{\'a}tica, FCEyN}
   \centerline{Universidad de Buenos Aires}
   \centerline{Pabell\'{o}n I, Ciudad Universitaria (1428),
Buenos Aires, Argentina}
} 

\medskip

\centerline{\scshape Nicol\'{a}s Frevenza}
\medskip
{\footnotesize
 \centerline{Departamento de M{\'e}todos Matem{\'a}ticos y Cuantitativos, FCEA}
   \centerline{Universidad de la Rep\'{u}blica}
   \centerline{Gonzalo Ram{\'i}rez 1926 (11200), Montevideo, Uruguay}
}

\medskip

\centerline{\scshape Julio D. Rossi}
\medskip
{\footnotesize
 \centerline{CONICET and Departamento  de Matem{\'a}tica, FCEyN}
   \centerline{Universidad de Buenos Aires}
   \centerline{Pabell\'{o}n I, Ciudad Universitaria (1428),
Buenos Aires, Argentina}
}

\bigskip

 \centerline{(Communicated by the associate editor name)}

\begin{abstract}
We study convex and quasiconvex functions on a metric graph.
Given a set of points in the metric graph, we consider the largest convex function below the prescribed datum.
We characterize this largest convex function as the unique largest viscosity subsolution to a simple differential equation, $u''=0$ on the edges, plus nonlinear transmission conditions at the vertices.
We also study the analogous problem for quasiconvex functions and obtain a characterization of the largest quasiconvex function that is below a given datum. 
\end{abstract}

\section{Introduction and description of the main results}

Our main goal in this paper is to study convex and quasiconvex functions on a metric graph. 

Let us start this introduction by recalling the well-known definitions of convexity and quasiconvexity in the Euclidean space. 
A function  $u\colon S\to \mathbb{R}$ defined on a convex subset $S\subset \mathbb{R}^N$ is called convex if for all $x,y\in S$ and any $\lambda \in [0,1],$ we have
\[
u(\lambda x+(1-\lambda )y)\leq \lambda u(x)+(1-\lambda )u(y).
\]
That is, the value of the function at a point in the segment that joins 
$x$ and $y$ is less or equal than the convex combination between 
the values at the extreme.
An alternative way of stating convexity is to say that $u$ is convex 
on $S$ if the epigraph of $u$ on $S$ is a convex set on $\mathbb{R}^{N+1}$. 
We refer to \cite{NicuPer, Vel} for general references on convex structures. 

A notion weaker than convexity is quasiconvexity. 
A function 
$u\colon S\to \mathbb{R} $ defined on a convex subset $S$ 
of the Euclidean space is called quasiconvex 
if for all $x,y\in S$ and any $\lambda \in [0,1],$ 
we have
\[
u(\lambda x+(1-\lambda )y)\leq \max \left\{ u(x),u(y)\right\}.
\]
An alternative and more geometrical way of defining a quasiconvex function 
$u$ is to require that each sublevel set $S_{\alpha }(u)=\{x\in S\colon u(x)\leq \alpha \}$
is a convex set. See \cite{10} and citations therein for an overview.

	Notice that whether
or not a function is convex depends on the numbers which the function assigns
to its level sets, not just on the shape of these level sets. The problem with this is that a
monotone transformation of a convex function need not be convex; that is,
if $u$ is convex and $g:\mathbb{R} \mapsto \mathbb{R}$ is increasing, then 
$g\circ u$ may fail to be convex. For instance, $f(x)=x^2$ is convex and 
$g(x)=\arctan(x)$ is increasing but $g\circ f(x)=\arctan(x^2)$ is not convex.  
However, the weaker condition, quasiconvexity, maintains this quality under monotonic
transformations. Moreover, every monotonic transformation of a convex function
is quasiconvex (although it is not true that every quasiconvex function
can be written as a monotonic transformation of a convex function). 

Convex and quasiconvex functions have applications in a wide range of disciplines, for example, mathematical analysis, 
optimization, game theory, and economics (see \cite{DiGuglielmo, Pearce, Ko, Sion, Vel}).

In the Euclidean space $\mathbb{R}^N$, 
there is also a Partial Differential Equation approach for convex and quasiconvex functions, 
see \cite{AO,BGJ12a,BGJ12b,BGJ13,BlancRossi,Ober33,OS}. In fact, a function $u$ in the Euclidean space is 
convex if and only if it is a viscosity sub-solution to 
\begin{equation}\label{ec-RN.convex}
\min_{v \colon |v|=1} 
\langle D^2 u(x) v, v \rangle = 0,
\end{equation}
(for a proof, see Theorem 2 in \cite{Ober33}) and is quasiconvex 
if and only if it is a viscosity sub-solution to
\begin{equation}\label{ec-RN}
 \min_{ \substack{v \colon |v|=1, \\
\langle v, \nabla u(x) \rangle =0}} 
\langle D^2 u(x) v, v \rangle = 0
\end{equation}
(now we refer to Section 2, Theorem 2.6 and
Theorem 2.7, in \cite{BGJ13}).
Moreover, the convex and the quasiconvex envelope of a boundary datum are solutions to \eqref{ec-RN.convex} and 
\eqref{ec-RN}, respectively. For numerical approximations we refer to \cite{AO2,AO}.

When one wants to expand the notion of convexity or quasiconvexity to an ambient space beyond
the Euclidean setting, the key is to introduce what is a segment
in our space.
For notions of convexity in discrete settings (like graphs and lattices) we refer 
to \cite{Ca,DPFRconvex,FaJa1,FaJa2,Mu1,Mu2,Mu3,Pel} and references therein.
For viscosity solutions to elliptic equations in finite graphs we refer to \cite{MOS} and for
nonlocal equations related to game theory to \cite{EB}.

\subsection{Metric graphs} \label{sect.prelim}

We start gathering some basic facts about metric graphs, see for instance \cite{BK} and references therein.

A graph $\Gamma$ consists of a finite set of vertices $\V(\Gamma)=\{\v_i\}$ and a set of edges $\E(\Gamma)=\{\e_j\}$ connecting some of the vertices. 
The graph $\Gamma$ is simple when there is not an edge connecting a vertex with itself.
A graph $\Gamma$ is said a finite graph if the number of edges and the number of vertices are finite.
Two vertices $\u$ and $\v$ are called adjacent (denoted $\u\sim \v$) if there is an edge connecting them. 
We denote the set of all vertices adjacent to $\v$ by $\V_\v(\Gamma).$ 
An edge $\e\in\E$ is incident to $\v\in \V$ when $\e$ connects $\v$ to another vertex and we denote it by $\e\sim \v$.
We define $\E_{\v}(\Gamma)$ as the set of all edges incident to $\v.$ 
The degree $d_{\v}(\Gamma)$ of a vertex $\v$ is the number of edges that incident to it. 
When there is no confusion, $\Gamma$ will be omitted from the notation.
A vertex $\v\in \V$ is called an interior vertex if $d_{\v}>1$.
Otherwise, we say that $\v$ is exterior. 
The set all interior (exterior) vertices is denoted by $\V_{int}$ ($\V_{ext}$). 
We will also refer to the exterior vertices as terminal vertices.

\begin{assu} 
\label{condicion1}
Throughout this article we assume that all graphs are connected, simple and with bounded degree, that is, $1\leq \sup_{\v} d_{\v}<\infty.$ 
\end{assu}

We consider also an orientation to each edge of $\Gamma$, that is, there is a map $\phi\colon \E\to \V\times\V$ associating to each edge $\e\in \E$ the pair $(\e_{-},\e_{+})\in \V\times\V$ of initial vertex and terminal vertex respectively. 
The edge $\hat{\e}$ is called the reversal of the edge $\e$ if $\hat{\e}_{-}=\e_{+}$ and $\hat{\e}_{+}=\e_{-}.$ 

\begin{definition}[See Definition 1.3.1 in \cite{BK}]
A graph $\Gamma=(\V,\E)$ with a map orientation $\phi$ is called a metric graph, if
\begin{enumerate}

\item Each edge $\e$ is assigned a positive length $\ell_{\e}\in(0,\infty].$
If $\ell_{\e}=\infty$, then $\e$ has only one vertex due to the other end goes to ``infinity".

\item The lengths of the edges that are reversals of each other are assumed to be equal, that is, $\ell_{\e}=\ell_{\hat{\e}};$

\item A coordinate $x_{\e}\in I_{\e}=[0,\ell_{\e}]$ is increasing in the direction of the edge given is assigned on each edge by $\phi$;

\item The relation $x_{\hat{\e}}=\ell_{\e}-x_{\e}$ holds between the coordinates on mutually reversed edges.
\end{enumerate}
\end{definition}
For an edge $\e$ with associated interval $[0,\ell_{\e}]$, the vertices $\e_{-}$ and $\e_{+}$ are identified with the coordinates $0$ and $\ell_{\e}$ respectively.
For a coordinate $x\in I_{\e}$ sometimes we write $x\in\e$.

A sequence of edges $\{\e_j\}_{j=1}^n\subset \E$ forms a path, and its length is defined as $\sum_{j=1}^n\ell_{\e_j}.$
Note that we are not considering the orientation map to define paths.
For two vertices $\v$ and $\u,$ the distance $d_0(\v,\u)$ is defined as the
length of the shortest path between them. 
When two points $x$ and $y$ are located at the same edge $\e$, that is, $x,y\in I_{\e}=[0,\ell_{\e}]$, the distance between them is defined by $d_{\e}(x,y) =  |x-y|$.
The distance $d$ in the metric graph $\Gamma=(\V,\E,\phi)$ is the natural extension of the previous defined distances, that is, 
\[
d(x,y) \coloneqq 
\inf 
\left\{ 
d_{\e}(x,z_1) + d_0(z_1,z_2) + d_{\bar{\e}}(z_2,y) 
\colon z_1,\,z_2\in \V, \, z_1\in \e,\, z_2\in \bar{\e}
\right\}
,
\]
where $x\in \e$ and $y\in \bar{\e}$ are two points that are not necessarily vertices or points at the same edge.
For $x,y\in\Gamma,$ we denote by $[x,y]$ the minimal path between $x$ and $y$. A metric graph $\Gamma$ becomes a metric measure space with the distance $d$ and the measure obtained from the standard Lebesgue measure on each edge.

The metric graph $\Gamma$ is connected and compact when it is connected and compact in the sense of a topological space.

\begin{assu}
\label{condicion2}
We assume that $\Gamma$ is a connected compact metric graph. 
We also assume that if $x,y\in [\e_-,\e_+]$ for some $\e\in\E$ then $d(x,y)=|x-y|.$
\end{assu}

A function $u$ on a metric graph $\Gamma$ is a collection of functions $u_{\e}$ defined on $[0,\ell_{\e}]$ for all $\e\in \E,$ not just at the vertices as in discrete models.
The space $C^k(\Gamma)$ consists of all continuous function on that belong to  $C^k(\e) \coloneqq C^k(I_\e)$ for each $\e\in\E.$
Let $u\in C^1(\Gamma),$ $\v\in\V$ and $\e\in\E_{\v}$, we define the ingoing derivative of $u$ over the edge $\e$ in $\v$ as follows
\[
\dfrac{\partial u}{\partial x_{\e}}(\v)
=
\left\{
\begin{array}{rl}
u_{\e}^\prime(\v) &\text{if } \v=\e_{-},\\
-u_{\e}^\prime(\v) &\text{if } \v=\e_{+}
\end{array}
\right.
,
\] 
that is, $\tfrac{\partial u}{\partial x_{\e}}(\v)$ is the
directional derivative taken in the direction into the edge starting at $\v.$

\subsection{Convex functions in metric graphs} 
We use the classical notion of convexity.
\begin{definition} 
\label{def.convex.intro}
A function $u\colon \Gamma \mapsto \mathbb{R}$ is \emph{convex} when for any $x,y\in \Gamma$ satisfies
\[
u(z) 
\leq 
\frac{d(y,z)}{d(x,y)} u(x)+ \frac{d(x,z)}{d(x,y)} u(y),
\]
for any $z \in[x,y]$.
\end{definition}

\begin{remark}
Note that convexity does not depend on the orientation map $\phi$ for the edges.
\end{remark}

As an example of a $C^2$ function that is convex in a star-shaped graph (a metric graph with only one node with 
multiplicity higher than one) we mention 
$u(x) = (d(x,x_0))^2$ with $x_0$ the unique multiple node of the graph. 

We are interested in the largest convex function that is below a given datum in some subset of the graph. 
Let $A\subset \Gamma$ be a closed set and $f\colon A \to\mathbb{R}$ a bounded function.
Then we define,		
\begin{equation} 
\label{convex-envelope}
u^*_f (x) 
\coloneqq 
\sup \left\{u(x) \colon u\in\mathcal{C}(f) \right\},
\end{equation}
where
$
\mathcal{C}(f) 
\coloneqq 
\left\{ 
u\colon \Gamma \to \mathbb{R}
\colon 
u \text{ is \emph{convex} and } 
u(x)\leq f(x), \ x\in A 
\right\}.
$

\begin{remark} \label{u*_f_bien_definida}
Observe that $\mathcal{C}(f)\neq \emptyset$ due to the fact that $u(x)\equiv \inf_{y\in A} \{f(y)\}$ is a convex function.
The function $u^*_f$ is well-defined and convex, since the supremum of convex functions is also a convex function.
\end{remark}

\begin{remark}\label{rem.hp} 
When the set $A$ is the whole graph, $A=\Gamma$, we have that $u^*_f $ is the usual convex envelope of $f$ in $\Gamma$. 
When $A$ is strictly contained in $\Gamma$ we have a convex envelope of $f$ extended as $+\infty$ to $\Gamma \setminus A$ (that is, we deal with the convex envelope of a partial datum).
Notice that it may happen that $u^*_f (x)< f(x)$ for some points $x \in A$ (it could be the case that there is no convex function that agrees with $f$ in $A$). 
When $u^*_f$ agrees with $f$ in the whole $A$ we have an optimal convex extension of $f$ to the set $\Gamma \setminus A$ (optimal in the sense of being the largest). 
\end{remark}

Our first result states when $u_f^*$ is bounded.

\begin{theorem} 
\label{teo-finita.intro}
Let $A\subset \Gamma$ be closed and $f\colon A \to \R$ bounded.
Consider $u^*_f$ be given by \eqref{convex-envelope}. 
Then, $u^*_f$ is bounded on $\Gamma$ if and only if $A$ contains every terminal node of $\Gamma$. 
In this case, we have
\[
\inf_A f 
\leq 
u^*_f (x) 
\leq  
\sup_A f 
\qquad \forall x \in \Gamma.
\]
\end{theorem}

Next, we show an equation together with a nonlinear coupling at the nodes that characterizes $u_f^*$ on $\Gamma$. 

Our first result is a characterization of convex functions in a metric graph.
\begin{theorem}\label{teo:convex-function.intro}
A function $u\colon\Gamma \mapsto \mathbb{R}$ is convex if and only if $u$ is a viscosity solution 
to
\begin{equation} \label{eq:ce.intro}
\begin{array}{rl}
u^{\prime\prime} \geq 0, 
\quad & \text{on the edges of } \Gamma \\[6pt]
\displaystyle \min_{\e,\bar{\e}\in \E_{\v}}
\left\{
\dfrac{\partial u}{\partial x_{\e}}(\v)
+
\dfrac{\partial u}{\partial x_{\bar{\e}}}(\v) 
\right\}
\geq 0, 
\quad &
\text{if } \v\in\V_{int}.
\end{array}
\end{equation}.
\end{theorem}

Next, we characterize the largest convex function below $f$ in $A$, $u^*_f$,  in terms of an obstacle problem.

\begin{theorem} 
\label{teo-ecuacion.intro}
Let  $u^*_f$ be given by \eqref{convex-envelope} for a given datum $f$ defined in $A\subset \Gamma$, 
where $f$ is bounded and $A$ closed. 
Let the contact set be given by
$$
C = \{ x\in A: u^*_f(x) = f(x) \}.
$$
Then, $u^*_f$ is a viscosity solution to
\begin{equation} \label{teo-ecuacion.intro.eq2}
\begin{array}{rl}
u''=0, & \text{ on the edges of } \Gamma\setminus C, \\[6pt]
\displaystyle \min_{\e, \bar{\e}\in \E_{\v}} 
\left\{
\dfrac{\partial u}{\partial x_{\e}}(\v) + \dfrac{\partial u}{\partial x_{\bar{\e}}}(\v)
\right\}
= 0, 
& \text{ for any node } \v\in \Gamma \setminus C,
\end{array}
\end{equation}
and therefore $u^*_f$ is the solution to the obstacle problem for the equations in \eqref{teo-ecuacion.intro.eq2}
with $f|_C$ as obstacle. 
\end{theorem}

\begin{remark}
Notice that the result covers the problem for the convex envelope (when $f$ is given in the whole graph $\Gamma$) and the optimal convex extension problem (the situation when there is a convex function that agrees with $f$ in $A$ and hence $u_f^*=f$ in $A$). 
\end{remark}

Notice that for a finite metric graph we have a finite number of degrees of freedom for the largest convex function below $f$ in $A$.
In fact, we have for each edge two degrees of freedom (since $u_f^*$ is a solution to $u''=0$ on each edge we have that it takes the form $u(x)=ax+b$). 
Therefore, to find $u_f^*$ we just have to select the constants $a$, $b$, on each edge such that the resulting function is continuous, verifies the nonlinear condition in \eqref{teo-ecuacion.intro.eq2} at the nodes and agrees with the given datum in $C$.

The equation \eqref{teo-ecuacion.intro.eq2} in the metric graph is the analogous to \eqref{ec-RN.convex} in the Euclidean space.
In fact, notice that on the edges there is only one direction (and the equation \eqref{ec-RN.convex} says that 
the second derivative in that direction is zero) 
and at a vertex the nonlinear condition says that in the union of two edges that contain the vertex (a direction)  
the second derivative of $u$ is zero while is greater or equal than zero in any other possible direction.

A quantum graph is a metric graph in which we associate a differential law with each edge
	with a coupling condition on the nodes, see \cite{BK}. Quantum graphs (in contrast to more elementary graph models, such as simple
unweighted or weighted graphs) are used to model
thin tubular structures, so-called graph-like spaces, they are their natural
limits, when the radius of a graph-like space tends to zero, see \cite{BK}.
Remark that our convex envelope is characterized 
	as being affine in each edge (a solution to the linear equation $u''=0$), and verifies a nonlinear condition 
	at the nodes (a $\min$ is involved). Therefore, the characterization of the convex envelope turns $\Gamma$ into
	a quantum graph.

\subsection{Quasiconvex functions in metric graphs} 
Now we turn out attention to quasiconvex functions in a metric graph $\Gamma$.
As for the convex case, let us use the classical definition.
 
\begin{definition} \label{def.quasiconvex.intro}
A function $u:\Gamma \mapsto \mathbb{R}$ is \emph{quasiconvex} if for any $x,y\in \Gamma$ we have
\[
u(z)
\leq
\max \{ u(x) ; u(y)\},
\]
for any $z\in [x,y].$ 
\end{definition}

For $A\subset \Gamma$ closed and $f\colon A \to\mathbb{R}$ bounded, the 
largest quasiconvex function on $\Gamma$ that is below $f$ in $A$ 
is defined as follows:
\begin{equation} 
\label{quasiconvex-envelope}
u^{\circledast}_f (x) 
\coloneqq 
\sup \left\{u(x) \colon u\in\mathcal{QC}(f) \right\},
\end{equation}
where
$
\mathcal{QC}(f) \coloneqq
\left\{ 
u\colon\Gamma \to\mathbb{R}
\colon 
u \text{ is \emph{quasiconvex} and } 
u(x)\leq f(x), \ x\in A 
\right\}. 
$
Observe that $\mathcal{QC}(f)\neq\emptyset$ since, for instance, $u^*_f\in \mathcal{QC}(f)$ (a convex function is also quasiconvex, so, $u^*_f$ is quasiconvex). 
Moreover, we have that  
\[ 
u_f^*(x)\le u^{\circledast}_f(x)
\qquad \text{for all } x\in\Gamma.
\] 

\begin{remark} A remark analogous to Remark \ref{rem.hp} is also useful here. 
When the set $A$ is the whole graph, $A=\Gamma$, we have that $u^{\circledast}_f$ is the quasiconvex envelope of $f$ in $\Gamma$. On the other hand, when $A$ is strictly contained in $\Gamma$, $u^{\circledast}_f$ is the quasiconvex envelope of $f$ extended by $+\infty$ to $\Gamma \setminus A$ (that is, we deal with the quasiconvex envelope of a partial datum).
Notice that it may happen that $u^{\circledast}_f (x)< f(x)$ for some points $x \in A$ (it could be the case that there is no quasiconvex function in $\Gamma$ that agrees with $f$ in $A$). 
When $u^{\circledast}_f$ agrees with $f$ in the whole $A$ we have an optimal quasiconvex extension of $f$ to the complement of $A$, $\Gamma \setminus A$ (optimal in the sense of being the largest). 
\end{remark}

For this optimal quasiconvex function $u^{\circledast}_f$, we have a discrete equation on the vertices, and in the edges, the function is piecewise constant. 
Notice that, in general, $u^{\circledast}_f$ is discontinuous.

\begin{theorem} 
\label{teo-quasiconvex.intro.88} 
Let $u^{\circledast}_f$ be given by \eqref{quasiconvex-envelope} for a given bounded datum $f$ defined in $A\subset \Gamma$, where $A$ is closed. 
Then it holds that $u^{\circledast}_f$ is bounded if and only if $A$ verifies that the convex hull of $A$ is the whole $\Gamma$, \emph{i.e.}, $\Co(A)= \Gamma$. 
In that case, we have
\[
\inf_A f \leq u^{\circledast}_f (x) \leq  \sup_A f \qquad \forall x \in \Gamma.
\]

Moreover, let the contact set be given by
$$
C = \{ x\in A: u^{\circledast}_f (x) = f(x) \}.
$$
then $u^{\circledast}_f$ verifies 
\begin{equation} \label{eq.cuasi}
\begin{array}{rl}
\displaystyle u(x) = \max \{u(\e_+);u(\e_-)\}, & \quad \text{if }x\in \e, \, \e\in \E \setminus C,\\[6pt]
\displaystyle u(\v) 
= 
\min\limits_{\substack{\u,\w\in \V_\v \\ \u\neq\w }} \max\{u(\u),u(\w)\} & \quad \text{if } \v\in\V\setminus C,
\end{array}
\end{equation}
where $\V_{\v}$ denotes the set of vertices that are adjacent to $\v$.
Therefore, $ u^{\circledast}_f$ is the solution to the obstacle problem for the equations in \eqref{eq.cuasi}
with $f|_C$ as obstacle.
\end{theorem}

As happens in the convex case, for the quasiconvex case, we have a finite number of degrees of freedom.
In fact, we have only to obtain the values of $u^{\circledast}_f$ at the vertices of $\Gamma$. 
The values at these points are uniquely determined by the relation 
$u^{\circledast}_f(\v) = \min_{\u,\w \in \V_{\v}} \max  \{u^{\circledast}_f(\u);u^{\circledast}_f(\v)\},$ 
that says that the value of $u^{\circledast}_f$ at $\v$ is the second one among the values at nodes that are adjacent to $\v$ (ordering these values from the smallest to the largest).

\bigskip

The paper is organized as follows: 
in Section \ref{sect.proofs.ce} we deal with the convex case; 
in Section \ref{sect-quasi} we prove our results for the quasiconvex case; 
and, finally, in Section \ref{sect-ejemplos} we collect some examples that illustrate our results.

\section{Convex functions} \label{sect.proofs.ce}
Let us start this section by recalling our definition of a convex function.
A function $u\colon \Gamma \mapsto \mathbb{R}$ is \emph{convex} if for any $x,y\in \Gamma$ we have
\[
u(z)\leq \frac{d(y,z)}{d(x,y)} u(x)+ \frac{d(x,z)}{d(x,y)} u(y) 
\qquad \text{for all } z\in [x,y]
.
\]

\begin{remark}
\label{re:regularity}
Let $u\colon \Gamma \mapsto \mathbb{R}$ be a convex function.
Using the smoothness properties of convex functions on intervals, we have that
\begin{itemize}
	\item $u$ is  upper semi-continuous  on $\Gamma$;
	
	\item $u$ is continuous on $\Gamma^\prime=\Gamma\setminus \V_{ext}.$
	In fact $u$ admits left and right derivatives on $\Gamma^\prime$, and these are monotonically non-decreasing. 
	As a consequence, $u$ is differentiable at all but at most countably many points on $\Gamma^\prime.$
	\item Finally, by Alexandrov's theorem, $u$ is almost everywhere twice differentiable on $\Gamma$. 
\end{itemize}

We refer to \cite{NicuPer} for the proofs of these facts.
\end{remark}

\begin{remark}
\label{re:constantes}
Keeping in mind that the only convex functions on a circle are the constants, we have that when $u$ is a convex function on $\Gamma,$ then $u$ is constant on every closed minimal path.
\end{remark}

Now, we need to introduce the notion of viscosity sub(super)-solution to the problem
\begin{equation} \label{eq:ce}
\begin{array}{rl}
u^{\prime\prime} = 0, 
\quad & \text{on the edges of } \Gamma \\[6pt] \displaystyle 
\min_{\e,\bar{\e}\in \E_{\v}}
\left\{
\dfrac{\partial u}{\partial x_{\e}}(\v)
+
\dfrac{\partial u}{\partial x_{\bar{\e}}}(\v) 
\right\}
=0, 
\quad &
\text{if } \v\in\V_{int},\\[6pt]
u(\v) =
\lim\limits_{x\to \v} u(x), 
\quad & \text{if } \v\in\V_{\normalfont ext}.
\end{array}
\end{equation}

\begin{definition}
Let $u\colon\Gamma\to\mathbb{R}$ be an upper (lower) semicontinuous function. 
We say that $u$ is a viscosity sub(super)-solution to \eqref{eq:ce} if only if 
\begin{itemize}

\item For every $x_0\in \Gamma\setminus \V$ and every time there exist $\delta>0$ and a test function $\varphi\in C^2(x_0-\delta,x_0+\delta)$ such that $\varphi(x_0)=u(x_0)$ and $\varphi(x)\ge u(x)$ ($\varphi(x)\le u(x)$) for all $x\in(x_0-\delta,x_0+\delta),$ then
\[
\varphi^{\prime\prime}(x_0) \ge 0 \, (\le 0);
\]

\item For every $\v\in \V_{int}$ and every time there exist $\e,\bar{\e}\in\E_{\v}$ and a test function $\varphi\in C^1(\e\cup\bar{\e})$ such that $\varphi(\v)=u(\v)$ and $\varphi(x)\ge u(x)$ ($\varphi(x)\le u(x)$) for all $x\in\e\cup\bar{\e},$ then
\[
\dfrac{\partial \varphi}{\partial x_{\e}}(\v)
+
\dfrac{\partial \varphi}{\partial x_{\bar{\e}}}(\v) 
\geq 0 \,(\leq 0).
\]
\end{itemize}

A viscosity solution of  \eqref{eq:ce} is a continuous function $u$ which is at the same time a sub-solution and super-solution.
\end{definition}

\begin{remark}
The boundary condition in \eqref{eq:ce} is a direct consequence of the regularity
results for convex functions on a metric graph stated in Remark \ref{re:regularity}. 
\end{remark}

Our first result is a characterization of convex functions in a metric graph.

\begin{theorem}\label{teo:convex-function}
A function $u\colon\Gamma \mapsto \mathbb{R}$ is convex if and only if $u$ is a viscosity sub-solution 
of \eqref{eq:ce}.
\end{theorem}

\begin{proof}[Proof of Theorem \ref{teo:convex-function}]
Let us start by assuming that $u$ is a convex function. 
If $x_0\in \Gamma\setminus \V$ 
and there exist $\delta>0$ and a test function 
$\varphi\in C^2(x_0-\delta,x_0+\delta)$ such that $\varphi(x_0)=u(x_0)$ and
$\varphi(x)\ge u(x)$  for all $x\in(x_0-\delta,x_0+\delta),$
then we have that
\[
	\varphi(x_0) = u(x_0)
	\leq \frac{1}{2} u(x_0+\epsilon)+ 
	\frac{1}{2} u(x_0-\epsilon) \leq  \frac{1}{2} 
	\varphi(x_0+\epsilon)+ \frac{1}{2} \varphi(x_0-\epsilon)			
\]
for any $0<\varepsilon<\delta,$ due to the fact that $u$ is convex.	
It follows that 
\[
	\varphi'' (x_0) \geq 0.
\]
Therefore, $u$ holds \eqref{eq:ce}. 
A similar argument shows that $u$ satisfies the boundary condition and so, is a viscosity sub-solution to \eqref{eq:ce}. 

\medskip

We now assume that $u$ is a viscosity sub-solution of \eqref{eq:ce}.
We argue by contradiction and assume that $u$ is not convex.
First, we suppose that $u$ is not convex at $z\in \e,$ that is, there exist $x,y\in \e$ such that
\begin{equation}
\label{eq:contra-edge}
u(z)
> 
\dfrac{d(y,z)}{d(y,x)}u(x)
+
\dfrac{d(x,z)}{d(y,x)}u(y).
\end{equation}
We use an idea from \cite[Theorem 1]{Ober33}.
Let $q$ be the parabola which interpolates $u$ on $I_{\e}$ at the points $x,y,z$. 
By \eqref{eq:contra-edge}, $q^{\prime\prime}<0$.
Since the function $u-q$ is lower semi-continuous, it has a maximum $M$ on the compact set $[x,y]$. 
If $M=0$, we have that $\varphi=q$ is a test function that contradicts the definition of sub-solution for $u$.
When $M>0$ is positive, and it is attached at $t_0$ in the interior of the set $[x,y]$, the function $\varphi = q + M$ is a test function for which $u$ does not satisfy the sub-solution condition at $t_0$.

Now, assume that the convexity does not hold on $\v\in \V_{\normalfont int}.$
Then, there exist $\e,\bar{\e}\in\E_{\v},$ $x\in\e$ and $y\in\bar{\e}$ such that
\begin{equation}
\label{eq:contra1}
u(z)
> 
\dfrac{d(y,z)}{d(y,x)}u(x)
+
\dfrac{d(x,z)}{d(y,x)}u(y)
\end{equation}
for any $z\in [x,\v]\cup[\v,y].$ 
Then, if we define
\[
\varphi_{\e}(z)
=\begin{cases}
\dfrac{u(x)-u(\v)}{d(x,\v)}d(z,\v) + u(\v)&\text{ if } \e_-=\v,\\[4pt]
\dfrac{u(\v)-u(x)}{d(x,\v)}d(z,\v) + u(x)&\text{ if } \e_+=\v,
\end{cases},
\]
\[
\varphi_{\bar{\e}}(z)
=\begin{cases}
\dfrac{u(y)-u(\v)}{d(y,\v)}d(z,\v) + u(\v)&\text{ if } \bar{\e}_-=\v,\\[4pt]
\dfrac{u(\v)-u(y)}{d(y,\v)}d(z,\v) + u(y)&\text{ if } \bar{\e}_+=\v,
\end{cases}
\]
and
\[
\varphi(z)=
\begin{cases}
\varphi_{e}(z) &\text{ if } z\in\e,\\
\varphi_{\bar{e}}(z) &\text{ if } z\in\bar{\e}, 
\end{cases}
\]
we get that $\varphi\in C^1(\e\cup\bar{\e}).$ 
As $u$ is convex on every edge, it holds that $u$ is below $\varphi$ on $[x,\v]$ and $[\v,y]$. 
Therefore we conclude that
\[
\varphi(z)
\geq u(z), 
\qquad\forall z\in [x,y].
\] 
Moreover, by \eqref{eq:contra1} we have
\begin{align*}
	\dfrac{\partial \varphi}{\partial x_{\e}}(\v)
	+\dfrac{\partial \varphi}{\partial x_{\bar{\e}} }(\v)
	&= \dfrac{u(x)-u(\v)}{d(x,\v)}+	\dfrac{u(y)-u(\v)}{d(y,\v)}\\
	&= \dfrac{d(y,\v)u(x)+d(x,\v)u(y) - d(x,y)u(\v) }{d(x,\v)d(y,\v)}\\
	&<0.
\end{align*}
This gives a contradiction with the fact that $u$ is a viscosity sub-solution to \eqref{eq:ce}.
\end{proof}

We are now in a position to study the largest convex function that is below of a given datum in some subset of the graph. 
Recall that for $A\subset \Gamma$ a closed set and $f\colon A \to\mathbb{R}$ a bounded function, the optimal
convex function on $\Gamma$ that is below $f$ in $A$ is defined by
\[	u^*_f (x) \coloneqq 
	\sup \left\{u(x) \colon u\in\mathcal{C}(f) \right\},
\]
	where $ \mathcal{C}(f) 
	\coloneqq
	\left\{ u\colon\Gamma \to\mathbb{R}\colon 
	u \text{ is \emph{convex} and } 
	u(x)\leq f(x), \ x\in A 
	\right\}. $

First, we just observe that $u^*_f $ is convex. 
The proof of this fact is immediate and included only for completeness. 

\begin{lemma}
Let $A\subset \Gamma$ be a closed set and $f\colon A \to\mathbb{R}$ be a bounded function.
Then $u^*_f$ is a convex function.
\end{lemma}

\begin{proof}
For any $u\in\mathcal{C}(f)$ and $x,y\in \Gamma$ we have
\[
u(z)
\leq 
\frac{d(y,z)}{d(x,y)} u(x)
+ 
\frac{d(x,z)}{d(x,y)} u(y)
\leq
\frac{d(y,z)}{d(x,y)} u^*_f(x) 
+ 
\frac{d(x,z)}{d(x,y)} u^*_f(y),
\]
for any $z\in [x,y],$ and taking supremum it follows that $u^*_f$ is a convex function.
\end{proof}

\begin{proposition}
\label{prop:convexA}
Let $A\subset \Gamma$ be a closed set and $f\colon A \to\mathbb{R}$ be a bounded function. 
Then $u^*_f$ is bounded in $\Gamma$ if and only if $A$ verifies that every terminal node is in $A$, that is, $\V_{ext}\subset A$.
\end{proposition}

\begin{proof}
First, we assume that there is a terminal node $\v\in \Gamma$ such that $\v\not\in A$. 
As $A$ is closed there is an interval in the edge that contains $\v$ as one of its endpoints such that $(b,\v]\subset \e \subset \Gamma$ and $(b,\v]\cap A =\emptyset$. 
Then, for any $n\in\mathbb{N}$ such that $n\ge	\inf \{f(y)\colon y\in A\},$ 
the function
\[
u(x)=
\begin{cases}
\displaystyle 
\inf_{y\in A} \{f(y)\} & \text{if } x\not\in (b,\v],\\[4pt]
\displaystyle 
n (x-b) + \inf_{y\in A} \{f(y)\} & \text{if } x\in (b,\v],
\end{cases}
\]
belongs to $\mathcal{C}(f).$ 
Therefore, $u^*_f$ is not bounded. 

Now, we assume that $A$ contains every terminal node. 
Suppose, arguing by contradiction, that $u^*_f$ is not bounded.
Recall that any convex function is continuous on $\Gamma' = \Gamma\setminus \V_{ext}$ and upper semi-continuous on $\Gamma$.
Hence, there is a convex function $u \in \mathcal{C}(f)$ and a point $x_0\in \Gamma$ such that,
\[
u(x_0) = \max_{y\in \Gamma'} u(y)
\coloneqq M > \sup_A f.
\]
Consider the set $\mathcal{M}=\{ x\in \Gamma \colon u(x) =M\}$.
Notice that $\mathcal{M} \subset \Gamma\setminus A \subset \Gamma'$.
Hence, $\mathcal{M}$ is closed since $u$ is continuous.
Given $x\in \mathcal{M}$, we have that for any segment $[a,b]$ such that $x\in [a,b]$, we must have $u\equiv M$ on the whole segment.
This shows that $\mathcal{M}$ is an open set.
Since $\Gamma$ is assumed to be connected, we get $\mathcal{M}=\Gamma$ which contradicts the fact that $A\neq \emptyset$.
\end{proof}

\begin{remark}
The previous proof and the Remark \ref{u*_f_bien_definida} prove that if $A$ contains every terminal node, we have
\[
\inf_A f \leq u^*_f (x) \leq  \sup_A f \qquad \text{for all } x \in \Gamma.
\]
\end{remark}

Throughout the rest of this section, we assume that $A$ is a closed subset of $\Gamma$ such that $A$ contains every terminal node. 
Consequently, $u^*_f $, the largest convex function below a finite datum in $A$, is bounded.

Now we prove our result concerning the equation verified by $u^*_f $.
Here we assume that the contact set $
C = \{ x\in A: u^*_f(x) = f(x) \}$ coincides with the whole $A$ (notice that the optimal function $u^*_g$ associated with
$g:=f|_C$ coincides with $u^*_f$). 

\begin{theorem}
The function $u^*_f$ is a viscosity solution to
\begin{align}
\label{eq:ceuf}
u^{\prime\prime} &= 0, 
\quad \text{on the edges of } \Gamma\setminus A, \\
\label{eq:bcuf}
\min_{\e,\bar{\e}\in \E_{\v}}
\left\{
\dfrac{\partial u}{\partial x_{\e}}(\v)
+
\dfrac{\partial u}{\partial x_{\bar{\e}}}(\v) 
\right\}
&=0, 
\quad 
\text{if } \v\in\V_{int}\setminus A.
\end{align}
\end{theorem}

\begin{proof}
Since $u^*_f$ is convex on $\Gamma,$ by Theorem \ref{teo:convex-function},
we have that $u^*_f$ is a viscosity sub-solution of \eqref{eq:ce}. 
Thus, we only need to show that $u^*_f$ is a viscosity super-solution of \eqref{eq:ceuf}-\eqref{eq:bcuf}.
Suppose, arguing by contradiction, that $u^*_f$ is not a super-solution of \eqref{eq:ceuf}-\eqref{eq:bcuf}. 
We have two possibilities:

\medskip

\noindent \emph{First case:}
There exist 
$x_0\in \Gamma\setminus (\V \cup A),$  
$\delta>0$ 
and a test function 
$\varphi\in C^2(x_0-\delta,x_0+\delta)$ 
such that 
$(x_0-\delta,x_0+\delta)\cap A=\emptyset,$
$\varphi(x_0)=u^*_f(x_0)$ 
and 
$\varphi(x)\le u^*_f(x)$ 
for all 
$x\in(x_0-\delta,x_0+\delta),$
such that
\[
\varphi'' (x_0) > 0.
\]
Since $\varphi\in C^2(x_0-\delta,x_0+\delta),$ without loss of generality,
we can assume that 
$\varphi'' (x) 
> 
\epsilon$ 
for all
$x\in(x_0-\delta,x_0+\delta)$
for $\varepsilon$ small enough.
Let $r$ be the tangent line to $\varphi$ at $x_0$,
\[
r(x)
\coloneqq 
\varphi(x_0) + \varphi'(x_0)(x-x_0),
\]
and define $\tilde{r}(x)
\coloneqq 
r(x)+\epsilon
.
$

Since $u_f^*$ is convex, there exist 
$z_1,z_2\in (x_0-\delta,x_0+\delta)$
such that 
$x_0\in[z_1,z_2],$ 
$u_f^*(z_1)=\tilde{r}(z_1),$ 
$u_f^*(z_2)=\tilde{r}(z_2),$
and 
$\tilde{r}(x)\ge u_f^*(x)$ 
for any $x\in[z_1,z_2].$
It follows that $w\colon \Gamma \to \mathbb{R}$ defined as
\[
w(x)=
\begin{cases}
\tilde{r}(x) & \text{if } x\in[z_1,z_2],\\
u_f^*(x) & \text{if } x\not\in[z_1,z_2],
\end{cases}
\] 
is a convex function and verifies 
\[
w(x_0) = \varphi(x_0) + \epsilon > u_f^*(x_0),
\]
with $w(x) = u_f^*(x) \leq f(x)$ if $x\in A.$
Then, it contradicts the the fact that $u_f^*$ is the supremum of convex functions below $f$ in $A$.
Therefore, $u^*_f$ is viscosity super-solution of \eqref{eq:ceuf}. 

\medskip

\noindent \emph{Second case:}  
There exists $\v\in \V_{int}\setminus A,$ such that for any $\e,\bar{\e}\in\E_{\v}$ for which there is a test function $\varphi\in C^1(\e\cup\bar{\e})$ so that $\varphi(\v)=u^*_f(\v)$ and $\varphi(x)\le u^*_f(x)$ for all $x\in \e\cup\bar{\e},$ we have 
\begin{equation}
\label{eq:2a1}
\dfrac{\partial \varphi}{\partial x_{\e}}(\v)
+
\dfrac{\partial \varphi}{\partial x_{\bar{\e}}}(\v) 
>0.
\end{equation}
Since $A$ is closed, there are $x\in\e,$ $y\in\bar{\e}$ such that
$[x,y]=[x,\v]\cup[\v,y],$ and $[x,y]\cap A=\emptyset.$ 

By the previous step, we have that $u_f^*$ is the viscosity solution of 
\[
u^{\prime\prime}=0 \quad\text{in } [x,\v]\cup[\v,y]\setminus\{\v\}
.
\]
Then $u_f^*$ is a linear function in $[x,\v]$ and $[\v,y].$ 
Hence, by \eqref{eq:2a1}
\begin{equation}
\label{eq:2a2}
\dfrac{\partial u^*_f}{\partial x_{\e}}(\v)
+
\dfrac{\partial u^*_f}{\partial x_{\bar{\e}}}(\v)
\geq
\dfrac{\partial \varphi}{\partial x_{\e}}(\v)
+
\dfrac{\partial \varphi}{\partial x_{\bar{\e}}}(\v) 
>0.
\end{equation}

Let $r$ be the linear function on $\e$ such that at the points $x$ and $\v$ reaches the values $u_f^*(x)$ and $u_f^*(\v)+\varepsilon$ respectively, where $\varepsilon>0$ will chosen later.
In analogous way define $\bar{r}$ linear on $\bar{\e}$ such that $\bar{r}(\v)=u_f^*(\v)+\varepsilon$ and $\bar{r}(y)=u_f^*(y).$

Using \eqref{eq:2a2} one can pick $\varepsilon>0$ small enough such that the sum of the ingoing derivatives at $\v$ of $r$ and $\bar{r}$ is still positive.
Hence, the function $w\colon \Gamma \to \mathbb{R}$ defined by
\[
w(z)
=
\begin{cases}
r(z) & \text{ for } z\in \e \\
\bar{r}(z) & \text{ for } z\in \bar{\e} \\
u_f^*(z) & \text{ otherwise}
\end{cases}
\]
is convex. 
We have also that $w(\v)=u_f^*(\v)+\varepsilon> u_f^*(\v)$ and $w$ restricted to $A$ is dominated by $f$, which is a contradiction with 
the definition of $u_f^*$. 
Therefore, $u^*_f$ is viscosity super-solution of \eqref{eq:bcuf}. 
\end{proof}

%
%
%
\section{Quasiconvex functions} \label{sect-quasi}

To begin this section, we recall the notion of quasiconvex function.
A function $u\colon \Gamma \mapsto \mathbb{R}$ is {\it quasiconvex}  if for any $x,y\in \Gamma$ we have
\[
u(z)\leq \max\{ u(x), u(y)\} \qquad \text{for any } z\in [x,y].
\]
%
%
%

Let $A\subset \Gamma$ be a closed set and $f\colon A \to\mathbb{R}$ be a bounded function. 
Recall that we want to study the largest quasiconvex function below $f$ in $A$ that is given by
\begin{equation} 
\label{quasiconvex-envelope.99}
u^{\circledast}_f (x) 
\coloneqq 
\sup \left\{u(x) \colon u\in\mathcal{QC}(f) \right\},
\end{equation}
where
$
\mathcal{QC}(f) \coloneqq
\left\{ 
u\colon\Gamma \to\mathbb{R}
\colon 
u \text{ is \emph{quasiconvex} and } 
u(x)\leq f(x), \ x\in A 
\right\}. 
$

Let us first prove that is quasiconvex as happens for convex functions.
Again the result is immediate, and we include the proof for completeness.

\begin{lemma}
Let $A\subset \Gamma$ be a closed set and $f\colon A \to\mathbb{R}$ be a bounded function.
Then $u^{\circledast}_f$ is a quasiconvex function.
\end{lemma}
\begin{proof}
For any $u\in\mathcal{QC}(f)$ and $x,y\in \Gamma$ 
we have
\[
u(z)
\leq 
\max
\{u(x),u(y)\}
\leq
\max\{u^{\circledast}_f(x),u^{\circledast}_f(y)\},
\]
for any $z\in [x,y].$ 
It follows that $u^{\circledast}_f$ is a quasiconvex function.
\end{proof}

Next, we turn our attention to the boundedness of $u^{\circledast}_f$. 
	
\begin{proposition}
\label{prop:quasiconvexA}
Let $A\subset \Gamma$ be a closed set and $f\colon A \to\mathbb{R}$ be a bounded function. 
Then, $u^{\circledast}_f$ is bounded if and only if $A$ verifies that $\Co(A)=\Gamma$.
\end{proposition}
	 	 
\begin{proof} 
First, assume that $u^{\circledast}_f$ is bounded. 
Arguing by contradiction, assume that $\Co(A) \neq \Gamma$ and consider the function
\[
u_n(x) = 
\begin{cases}
\inf\limits_A f & \text{for } x\in \Co(A), \\
n & \text{for }x\notin \Co(A).
\end{cases}
\]
The function $u_n$ is quasiconvex for every $n > \inf_A f $. 
Indeed, the sub level sets are 
\[
S_{\alpha }(u_n)
=
\{x\in \Gamma \colon u_n(x)\leq \alpha \} =
\begin{cases}
\emptyset & \text{for }\alpha < \inf\limits_A f, \\
\Co (A) & \text{for } \alpha \in [\inf\limits_A f , n),\\
\Gamma & \text{for } \alpha \geq n,
\end{cases}
\]
that are all convex subsets of $\Gamma$.
Since $n$ can be arbitrarily large, this contradicts that $u^{\circledast}_f$ is bounded. 

\medskip

Now, we assume that $A$ verifies that $\Co(A) =\Gamma$.
Let $x\in \Gamma$ be any point such that there are $a_1,a_2 \in A$ with $x \in [a_1,a_2]$. 
From the fact that $u^{\circledast}_f$ is quasiconvex we get that
\[
 u^{\circledast}_f(x) \leq \max \{f(a_1),f(a_2)\} \leq \sup_A f < \infty,
\]
proving that $u^{\circledast}_f$ is bounded in the set of convex combinations of points in $A$.
Then, we obtain that $u^{\circledast}_f$ is bounded by $\sup_A f $ in $\Co(A) =\Gamma$ which it concludes the proof.
\end{proof}

For $u^{\circledast}_f$ we have a discrete equation on the vertices, and in the edges, the function is piecewise constant.
Therefore, $u^{\circledast}_f$ is discontinuous in general.

As we did for convex functions, it can be proved that a function is quasiconvex if and only if it is a viscosity solution to
\begin{align}
u(x) 
& \leq
\max \{u(\e_+),u(\e_-)\}, \quad \text{for } x\in \e,\, \e\in\E, \\
u(\v) 
&  \leq 
\min\limits_{\substack{\u,\w\in \V_\v \\ \u\neq\w }} \max\{u(\u),u(\w)\}  \quad \text{for } \v\in\V\setminus A,
\end{align}
where $\V_{\v}$ denotes the set of vertices that are adjacent to $\v$. The proof is analogous to the one of Theorem
\ref{teo:convex-function} and therefore we omit it. 

Now we prove our result concerning the equation verified by $u^{\circledast}_f$.
As before, we assume that the contact set $C = \{ x\in A: u^{\circledast}_f(x) = f(x) \}$ coincides with the whole $A$ (notice that
$u^{\circledast}_g$  the largest quasiconvex function below 
$g:=f|_C$ in $C$ coincides with $u^{\circledast}_f$).

\begin{theorem} 
\label{teo-quasiconvex.intro} 
Consider a closed set $A\subset \Gamma$ such that $\Co(A)=\Gamma$ and let $f\colon A \to\mathbb{R}$ be a bounded function. 
Then, $u^{\circledast}_f$ verifies
\begin{align}
u(x) 
&= 
\max \{u(\e_+),u(\e_-)\}, \quad \text{for } x\in \e\setminus A,\, \e\in\E, \\
u(\v) 
&= 
\min\limits_{\substack{\u,\w\in \V_\v \\ \u\neq\w }} \max\{u(\u),u(\w)\}  \quad \text{for } \v\in\V\setminus A,
\end{align}
where $\V_{\v}$ denotes the set of vertices that are adjacent to $\v$.
\end{theorem}

\begin{proof}
We define the function $w\colon \Gamma \to \mathbb{R}$ as
\[
w(x)
\coloneqq
\left\{
\begin{array}{ll}
u^{\circledast}_f(x) 
& \text{if } x\in A, \\[4pt]
\max 
\{u^{\circledast}_f(\e_+),u^{\circledast}_f(\e_-)\} 
& \text{if } x\in \e\setminus A, \, \e\in\E, \\ [4pt]
\min\limits_{\substack{\u,\w\in \V_\v \\ \u\neq\w}}
\max\{u^{\circledast}_f(\u),u^{\circledast}_f(\w)\}
& \text{if } x=\v\in\V\setminus A.
\end{array}
\right. 
\]
Using that $u^{\circledast}_f$ is quasiconvex, it is easy to check that
$w\ge u^{\circledast}_f$ in $\Gamma.$ 
Moreover, since $u^{\circledast}_f$ is given by \eqref{quasiconvex-envelope.99}, we also have $w\ge f$ in $A.$ 
Therefore, to conclude the proof, it is enough to show that $w$ is a quasiconvex function.

Let $x,y\in \Gamma$ and $z\in [x,y].$ 
We split the rest of the proof into five cases:

\noindent \emph{Case 1:} 
$x,y,z\not\in\V$ and there is $\e\in\E$ such that $z\in\e$ and either $x\in\e$ or else $y\in \e.$

In this case, either $w(z)=w(x)$ or else $w(z)=w(y).$ Therefore
\[
w(z)\le\max\{w(x),w(y)\}
.
\]

\medskip

\noindent\emph{Case 2:} 
$z\not\in\V$ and there is $\e\in\E$ such that $z\in\e$ and either $x$ is vertex of $\e$ or else $y$ is vertex of $\e.$

In that case, either 
\[
w(z)\le \max\{u^{\circledast}_f(x),u^{\circledast}_f(\v)\},
\] 
or else 
\[
w(z)\le \max\{u^{\circledast}_f(y),u^{\circledast}_f(\v)\},
\] 
where here $\v$ denotes the other vertex of $\e.$ 
Since, $u^{\circledast}_f$ is quasiconvex and $\v\in[x,y]$ we have that
\[
u^{\circledast}_f(\v)\le \max\{u^{\circledast}_f(x),u^{\circledast}_f(y)\}
\le \max\{w(x),w(y)\},
\]
and therefore
$w(z)\le\max\{w(x),w(y)\}.$

\medskip

\noindent \emph{Case 3:}
$z\not\in\V,$ and there are $\v_1,\dots,\v_n\in\V$ such that
\[
[x,y]=[x,\v_1]\cup[\v_1,\v_2]\cup\cdots\cup[\v_n, y], 
\]
and $z\in[\v_j,\v_{j+1}]$ for some $j\in\{2,\dots,n-1\}.$ 

Then
\[
w(z)
\le 
\max
\{
u^{\circledast}_f(\v_j),u^{\circledast}_f(\v_{j+1})
\}
.
\]
Using that $u^{\circledast}_f$ is quasiconvex, we get
\[
\max\{
u^{\circledast}_f(\v_j),u^{\circledast}_f(\v_{j+1})
\}
\le 
\max\{
u^{\circledast}_f(x),u^{\circledast}_f(y)
\}
\le 
\max\{
w(x),w(y)
\}
.
\]
Therefore
$w(z)\le\max\{w(x),w(y)\}.$

\medskip

\noindent \emph{Case 4:}  
$z\in\V,$ there are $\e,\bar{\e} \in \E_{z}$ such that $x\in[\e_-,\e_+]$ and $y\in[\bar{\e}_-,\bar{\e}_+].$ 

Observe that
\[
w(x)
\ge 
u^{\circledast}_f(\v) 
\text{ and } 
w(y) 
\ge 
u^{\circledast}_f(\bar{\v}) 
\] 
where $\v$ and $\bar{\v}$ are the other vertices of $\e$ and $\bar{\e}$ respectively.
Then
\[
\max \{w(x),w(y)\}
\ge
\max\{u^{\circledast}_f(\v),u^{\circledast}_f(\bar{\v})\}
\ge w(z)
\]
due to $\v,\bar{\v}\in\V_z.$

\medskip

\noindent \emph{Case 5:}  
$z\in\V,$ and there are $\v_1,\dots,\v_{j-1},\v_{j+1},\dots,\v_n\in\V$ such that 
\[
[x,y] =
[x,\v_1]\cup\cdots\cup[\v_{j-1},z]\cup[z,\v_{j+1}]\cup\cdots \cup[\v_n, y].
\] 

Then,
\[
w(z)\le \max\{u^{\circledast}_f(\v_{j-1}),u^{\circledast}_f(\v_{j+1})\}.
\]
Using that $u^{\circledast}_f$ is quasiconvex, we get
\[
\max\{u^{\circledast}_f(\v_{j-1}),u^{\circledast}_f(\v_{j+1})\}
\le 
\max\{u^{\circledast}_f(x),u^{\circledast}_f(y)\}
\le 
\max\{w(x),w(y)\}.
\]
Therefore
$w(z)\le\max\{w(x),w(y)\}.$
\end{proof}

\section{Examples} \label{sect-ejemplos}

To illustrate our results, we include some examples. 
Recall that the convex and quasiconvex optimal functions do not depend on the orientation of the edges. 
However, we use the orientation of the edges to parametrize them and then 
describe a function on the metric graph $\Gamma$. 

In all of our examples we will assume that all edges have the same length $\ell_{\e}=1$ and 
are parameterized by (0,1) with the given orientation in the figure.

\begin{ex}
First, we give an example of a set $A$ that contains every terminal node, but the convex hull of $A$ is not the whole $\Gamma$. 
Consider the following graph,
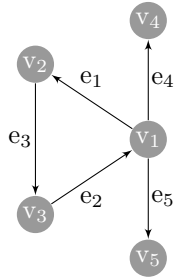
\begin{center}
\begin{tikzpicture}[rotate=-90,main_node/.style={circle,fill=black!40,minimum size=1em,inner sep=1pt]},scale=1]
\tikzset{edge/.style = {->,> = latex'}}
\node[main_node] (1) at (0,0) {\color{white}$\v_1$};
\node[main_node] (2) at (-1, -1.5) {\color{white}$\v_2$};
\node[main_node] (3) at (1, -1.5){\color{white}$\v_3$};
\node[main_node] (4) at (-1.58113883008,0) {\color{white}$\v_4$};
\node[main_node] (5) at (1.58113883008,0) {\color{white}$\v_5$};
\draw[edge] (1) to (2);
\draw[edge] (2) to (3);
\draw[edge] (3) to (1);
\draw[edge] (1) to (4);
\draw[edge] (1) to (5);
\node (6) at (-.8,-.75) {$\e_1$};
\node (7) at (.8,-.75) {$\e_2$};
\node (8) at (0,-1.7) {$\e_3$};
\node (9) at (-0.79056941504,0.2) {$\e_4$};
\node (10) at (0.79056941504,0.2) {$\e_5$};
\end{tikzpicture}
\end{center}

Set $A=\{\v_4,\v_5\}.$
Notice that $\Gamma$ has exactly two terminal nodes $\v_4$ and $\v_5$ (and $A$ is chosen precisely as the set of terminal nodes). 
However, the convex hull of $A$ is given by 
\[
\Co(A) = \{\v_1,\v_4, \v_5\} \cup \{\e_4,\e_5\} \neq \Gamma.
\]
In this example, if we set $f(\v_4) = a,$ $f(\v_5)=b,$ the function $u^*_f$ is given by
\[
u^*_f (x) = 
\left\{
\begin{array}{ll}
\tfrac{a+b}{2} & \text{ if } x\in \{\e_1,\e_2,\e_3\} \cup \{\v_1,\v_2,\v_3\},\\[4pt]
\tfrac{a+b}{2} + \tfrac{a-b}{2}x & \text{ if } x\in \e_4, \\[4pt]
\tfrac{a+b}{2} + \tfrac{b-a}{2}x & \text{ if } x\in \e_5, \\[4pt]
f(x) & \text{ if } x\in A.
\end{array}
\right.
\]

Also, in this example, the function
\[
u_n(x) = 
\left\{
\begin{array}{ll}
\min\{a,b\} & \quad x \in \Co(A), \\
n & \quad x \in \Gamma \setminus \Co(A),
\end{array}
\right.
\]
is quasiconvex for every $n > \min\{a,b\}$. 
Indeed, the sub level sets are 
\[
S_{\alpha }(u_n)=\{x\in \Gamma \colon u_n(x)\leq \alpha \} =
\left\{
\begin{array}{ll}
\emptyset & \text{ if } \alpha < \min\{a,b\}, \\
\Co (A) & \text{ if }\alpha \in [\min\{a,b\}, n), \\
\Gamma & \text{ if } \alpha \geq n,
\end{array}
\right.
\]
that are all convex subsets of $\Gamma$.
Since $n$ can be very large, we obtain that the supremum of quasiconvex functions below any datum on $A$ is not bounded.
\end{ex}

\begin{ex}
Consider the metric graph $\Gamma$ given in the following figure
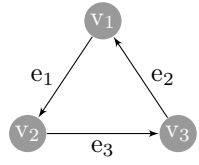
\begin{center}
\begin{tikzpicture}[main_node/.style={circle,fill=black!40,minimum size=1em,inner sep=1pt]},scale=1]
\tikzset{edge/.style = {->,> = latex'}}
\node[main_node] (1) at (0,0) {\color{white}$\v_1$};
\node[main_node] (2) at (-1, -1.5) {\color{white}$\v_2$};
\node[main_node] (3) at (1, -1.5){\color{white}$\v_3$};
\draw[edge] (1) to (2);
\draw[edge] (2) to (3);
\draw[edge] (3) to (1);
\node (4) at (-.8,-.75) {$\e_1$};
\node (5) at (.8,-.75) {$\e_2$};
\node (6) at (0,-1.7) {$\e_3$};
\end{tikzpicture}
\end{center}

Notice that in this example, there a no terminal nodes. If we fix just one value, say $f(\v_1)=a,$ we get that the 
largest convex function below $f$ at $\v_1$ is constant by Remark \ref{re:constantes}, $ u^*_f \equiv a$, while the 
corresponding largest quasiconvex function is not bounded.  

Now, consider three values at the nodes $f(\v_1)=a,$ $f(\v_2)=b$, $f(\v_3)=c,$ that is, $A=\{\v_1,v_2,\v_3\}.$
Then, $u^*_f$ is given by
\[
u^*_f(x) = 
\left\{
\begin{array}{ll}
a+(b-a)x  & \text{ if } x\in \e_1,\\[4pt]
c+(a-c)x  & \text{ if } x\in \e_2,\\[4pt]
b+(c-b)x  & \text{ if } x\in \e_3,\\[4pt]
f(x) & \text{ if } x\in \{\v_1,\v_2,\v_3\}.
\end{array}
\right. 
\]
Observe that $u^*_f$ is just the line that connects the boundary values on every edge. 

For the quasiconvex case, looking for $u^{\circledast}_f$, we assume that the given values are ordered as $a<b<c$.
Then, we have
\[
u^{\circledast}_f (x) =
\left\{
\begin{array}{rl}
c  & \text{ if } x\in \e_2\cup \e_3 \cup\{\v_3\},\\
b  & \text{ if } x\in \e_1\cup\{\v_2\},\\
a  & \text{ if } x=\v_1.
\end{array}
\right. 
\]

This example can be generalized to the circular graph with $n$ vertices $\v_1,\dots,\v_n$ and edges of the same length between $\v_i$ and $\v_{i+1}$ ($i=1,\dots,n-1$) and between $\v_1$ and $\v_n$.
\end{ex}

\begin{ex}
Let us consider a star-shaped graph as shown in the following figure,
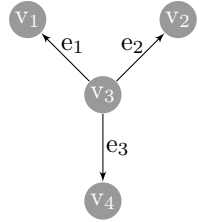
\begin{center}
\begin{tikzpicture}[main_node/.style={circle,fill=black!40,minimum size=1em,inner sep=1pt]}]
\tikzset{edge/.style = {->,> = latex'}}
\node[main_node] (3) at (0,0) {\color{white}$\v_3$};
\node[main_node] (2) at (1, 1) {\color{white}$\v_2$};
\node[main_node] (1) at (-1, 1){\color{white}$\v_1$};
\node[main_node] (4) at (0, -1.41421356237){\color{white}$\v_4$};
\draw[edge] (3) to (1);
\draw[edge] (3) to (2);
\draw[edge] (3) to (4);
\node (5) at (.4,.65) {$\e_2$};
\node (6) at (-.4,0.65) {$\e_1$};
\node (7) at (0.2,-0.70710678118) {$\e_3$};
\end{tikzpicture}
\end{center}
Then, if for instance, we set $A=\{\v_1,\v_2,\v_4\}$ and $f(\v_1)=0,$ $f(\v_2)=1$ and $f(\v_4)=2$, we have that
\[
u_f^*(x)=
\left\{
\begin{array}{ll}
\tfrac12 - \tfrac12 x & \text{ if } x\in\e_1,\\[4pt]
\tfrac12 + \tfrac12 x &\text{ if } x\in\e_2,\\[4pt]
\tfrac12+\tfrac32 x &\text{ if } x\in\e_3,\\[4pt]
\tfrac12  &\text{ if } x=\v_3,\\[4pt]
f(x)  &\text{ if } x\in A.
\end{array}
\right.
\text{ and }
%
u^{\circledast}_f(x)=
\left\{
\begin{array}{ll}	
1 & \text{ if } x\in\e_1 \cup \e_2  \cup \{\v_3\},\\[4pt]
2 & \text{ if } x \in \e_3,\\[4pt]
f(x) & \text{ if } x\in A.
\end{array}
\right.
\]

This example can be generalized to a star-shaped graph with $n+1$ vertices $\v_1,\dots,\v_{n+1}$ and edges of the same length between $\v_i$ and $\v_{n+1}$ for $i=1,\dots,n$.
\end{ex}

\begin{ex}
We consider the graph
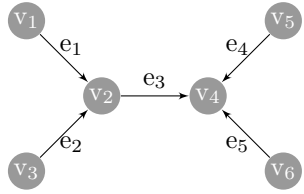
\begin{center}
\begin{tikzpicture}[main_node/.style={circle,fill=black!40,minimum size=1em,inner sep=1pt]}]
\tikzset{edge/.style = {->,> = latex'}}
\node[main_node] (1) at (-1, 1){\color{white}$\v_1$};
\node[main_node] (2) at (0,0) {\color{white}$\v_2$};
\node[main_node] (3) at (-1, -1){\color{white}$\v_3$};
\node[main_node] (4) at (1.41421356237, 0) {\color{white}$\v_4$};
\node[main_node] (5) at (2.41421356237, 1){\color{white}$\v_5$};
\node[main_node] (6) at (2.41421356237, -1){\color{white}$\v_6$};
\draw[edge] (1) -- (2);
\draw[edge] (3) -- (2);
\draw[edge] (2) -- (4);
\draw[edge] (5) -- (4);
\draw[edge] (6) -- (4);
\node (7) at (-.4,-.65) {$\e_2$};
\node (8) at (-.4,0.65) {$\e_1$};
\node (9) at (0.70710678118,0.2) {$\e_3$};
\node (10) at (1.80710678118,-.65) {$\e_5$};
\node (11) at (1.80710678118,0.65) {$\e_4$};
\end{tikzpicture}
\end{center}
Let $A=\{\v_1,\v_3,\v_5,\v_6\}$ and $f(\v_1)=0,$ $f(\v_3)=2,$ $f(\v_5)=1,$ $f(\v_6)=3.$
Then, 
\[
u_f^*(x)=
\left\{
\begin{array}{ll}
\tfrac13 x  & \text{ if } x\in\e_1,\\[4pt]
2 -\tfrac53 x & \text{ if } x\in\e_2,\\[4pt]
\tfrac13 + \tfrac13 x & \text{ if } x\in\e_3,\\[4pt]
1 -\tfrac13 x &\text{ if } x\in\e_4,\\[4pt]
3 -\tfrac73 x &\text{ if } x\in\e_5,\\[4pt]
\tfrac13  &\text{ if } x=\v_2,\\[4pt]
\tfrac23  &\text{ if } x=\v_4,\\[4pt]
f(x)  &\text{ if } x\in A,
\end{array}
\right.
\,
u^{\circledast}_f(x)=
\left\{
\begin{array}{cl}
1 &\text{if } x\in \e_1 \cup \e_3\cup e_4\cup \{ \v_2,\v_4\},\\[4pt]
2 &\text{if } x\in\e_2,\\[4pt]
3 &\text{if } x\in\e_5,\\[4pt]
f(x)  &\text{if } x\in A.
\end{array}
\right.
\]
\end{ex}

\begin{ex}
Finally, we consider a binary tree of two generations where edges are oriented to the root 
(see the figure below).
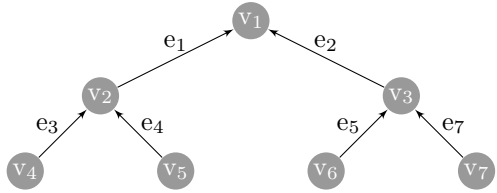
\begin{center}
\begin{tikzpicture}[main_node/.style={circle,fill=black!40,minimum size=1em,inner sep=1pt]}]
\tikzset{edge/.style = {->,> = latex'}}
\node[main_node] (1) at (0, 0){\color{white}$\v_1$};
\node[main_node] (2) at (-2,-1) {\color{white}$\v_2$};
\node[main_node] (3) at (2, -1){\color{white}$\v_3$};
\node[main_node] (4) at (-3, -2) {\color{white}$\v_4$};
\node[main_node] (5) at (-1, -2){\color{white}$\v_5$};
\node[main_node] (6) at (1, -2){\color{white}$\v_6$};
\node[main_node] (7) at (3, -2){\color{white}$\v_7$};
\draw[edge] (2) -- (1);
\draw[edge] (3) -- (1);
\draw[edge] (4) -- (2);
\draw[edge] (5) -- (2);
\draw[edge] (6) -- (3);
\draw[edge] (7) -- (3);
\node (8) at (-1,-.3) {$\e_1$};
\node (9) at (1,-.3) {$\e_2$};
\node (10) at (-2.7,-1.4) {$\e_3$};
\node (11) at (-1.3,-1.4) {$\e_4$};
\node (12) at (1.3,-1.4) {$\e_5$};;
\node (10) at (2.7,-1.4) {$\e_7$};
\end{tikzpicture}
\end{center}
Let $A=\{\v_4,\v_5,\v_6,\v_7\}$ and $f(\v_4)=0,$ $f(\v_5)=1,$ $f(\v_6)=2,$ 
$f(\v_7)=3.$
In this case, we have that
\[
u_f^*(x)=
\left\{
\begin{array}{ll}
\tfrac12 + \tfrac12 x & \text{ if } x\in\e_1,\\[4pt]
\tfrac32 - \tfrac12 x & \text{ if } x\in\e_2,\\[4pt]
\tfrac12 x & \text{ if } x\in\e_3,\\[4pt]
1- \tfrac12 x & \text{ if } x\in\e_4,\\[4pt]
2-\tfrac12 x & \text{ if } x\in\e_5,\\[4pt]
3- \tfrac32 x & \text{ if } x\in\e_6,\\[4pt]
1 & \text{ if } x=\v_1,\\[4pt]
\tfrac12 & \text{ if } x=\v_2,\\[4pt]
\tfrac32 & \text{ if } x=\v_3,\\[4pt]
f(x) & \text{ if } x\in A,
\end{array}
\right.
\,
u^{\circledast}_f(x)=
\left\{
\begin{array}{cl}
1 & \text{if } x\in \e_3 \cup \e_4 \cup \{\v_2\},\\[4pt]
2 & \text{if } x\in \e_1 \cup \e_2 \cup \e_5 \cup \{\v_1,\v_3\},\\[4pt]
3 & \text{if } x\in\e_7,\\[4pt]
f(x) & \text{if } x\in A.
\end{array}
\right.
\]

This analysis can be extended to larger trees.
 
\end{ex}



%


\begin{thebibliography}{99}

\bibitem{EB} 
\newblock A. Elmoataz and P. Buyssens, 
\newblock On the connection between tug-of-war games and nonlocal PDEs on graphs, 
\newblock \emph{C. R. Mecanique}, \textbf{345}, (2017), 177--183.

\bibitem{AO2} 
\newblock B. Abbasi and A. M. Oberman, 
\newblock Computing the Level Set Convex Hull,
\newblock \emph{J. Scientific Computing}, \textbf{75}, (2018), 26--42.

\bibitem{AO} 
\newblock B. Abbasi and A. M. Oberman,  
\newblock A partial differential equation for the uniformly quasiconvex envelope, 
\newblock \emph{IMA J. Numer. Anal.}, \textbf{39}, (2019), 141--166.


\bibitem{BGJ12a}
\newblock E. N. Barron, R. Goebel, and R. R. Jensen,
\newblock Functions which are quasiconvex under linear perturbations,
\newblock {\em SIAM J. Optim.}, \textbf{22}(3), (2012), 1089--1108.

\bibitem{BGJ13}
\newblock E. N. Barron, R. Goebel, and R. R. Jensen,
\newblock Quasiconvex functions and nonlinear {PDE}s,
\newblock {\em Trans. Amer. Math. Soc.}, \textbf{365}(8), (2013), 4229--4255.

\bibitem{BGJ12b}
\newblock  E. N. Barron, R. Goebel, and R. R. Jensen,
\newblock The quasiconvex envelope through first-order partial differential equations which characterize quasiconvexity of nonsmooth functions,
\newblock {\em Discrete Contin. Dyn. Syst. Ser. B}, \textbf{17}(6), (2012), 1693--1706.

\bibitem{BK}
\newblock  G. Berkolaiko and P. Kuchment,
\newblock {\em Introduction to quantum graphs}, volume 186 of {\em Mathematical Surveys and Monographs},
\newblock American Mathematical Society, Providence, RI, 2013.

\bibitem{BlancRossi}
\newblock P. Blanc and J. D. Rossi,
\newblock Games for eigenvalues of the {H}essian and concave/convex envelopes,
\newblock {\em J. Math. Pures Appl.}, \textbf{127}(9), (2019), 192--215.

\bibitem{Ca}
\newblock J. C\'{a}ceres, A. M\'{a}rquez, and M. L. Puertas,
\newblock Steiner distance and convexity in graphs,
\newblock {\em European J. Combin.}, \textbf{29}(3), (2008), 726--736.

\bibitem{DPFRconvex}
\newblock L. M. Del Pezzo, N. Frevenza, and J. D. Rossi,
\newblock Convex envelopes on trees,
\newblock {\em J. Convex Anal.}, \textbf{27}(4), (2020), 1195--1218.

\bibitem{DiGuglielmo}
\newblock F. Di Guglielmo,
\newblock Nonconvex duality in multiobjective optimization,
\newblock {\em Math. Oper. Res.}, \textbf{2}(3), (1977), 285--291.

\bibitem{10}
\newblock A. Eberhard and C. E. M. Pearce,
\newblock Class-inclusion properties for convex functions,
\newblock in {\em Progress in optimization ({P}erth, 1998)}, volume 39 of {\em  Appl. Optim.}, Kluwer Acad. Publ., (2000), 129--133.

\bibitem{FaJa1}
\newblock M. Farber and R. E. Jamison,
\newblock Convexity in graphs and hypergraphs,
\newblock {\em SIAM J. Algebraic Discrete Methods}, \textbf{7}(3), (1986), 433--444.

\bibitem{FaJa2}
\newblock M. Farber and R. E. Jamison,
\newblock On local convexity in graphs,
\newblock {\em Discrete Math.}, \textbf{66}(3), (1987), 231--247.

\bibitem{Ko}
\newblock H. Komiya,
\newblock Elementary proof for {S}ion's minimax theorem,
\newblock {\em Kodai Math. J.}, \textbf{11}(1), (1988), 5--7.

\bibitem{MOS} J. Manfredi, A. Oberman and A. Sviridov,
\newblock Nonlinear elliptic Partial Differential Equations and $p-$harmonic 
functions on graphs,
\newblock {\em Diff. Integral Eq.},  \textbf{28}(1-2), (2015), 79--102.

\bibitem{Mu1}
\newblock K. Murota,
\newblock Discrete convex analysis,
\newblock {\em Math. Programming}, \textbf{83} (3, Ser. A), (1998), 313--371.

\bibitem{Mu2}
\newblock K. Murota,
\newblock {\em Discrete convex analysis},
\newblock SIAM Monographs on Discrete Mathematics and Applications, Society for Industrial and Applied Mathematics (SIAM), Philadelphia, PA, 2003.

\bibitem{Mu3}
\newblock K. Murota,
\newblock Recent developments in discrete convex analysis,
\newblock in {\em Research trends in combinatorial optimization}, Springer, Berlin, (2009), 219--260.

\bibitem{NicuPer}
\newblock C. P. Niculescu and L .E. Persson,
\newblock {\em Convex functions and their applications}, volume 23 of {\em CMS Books in Mathematics/Ouvrages de Math\'{e}matiques de la SMC}, Springer, New York, 2006.


\bibitem{Ober33}
\newblock A. M. Oberman,
\newblock The convex envelope is the solution of a nonlinear obstacle problem,
\newblock {\em Proc. Amer. Math. Soc.}, \textbf{135}(6), (2007), 1689--1694.


\bibitem{OS}
\newblock A. M. Oberman and L. Silvestre,
\newblock The {D}irichlet problem for the convex envelope,
\newblock {\em Trans. Amer. Math. Soc.}, \textbf{363}(11), (2011), 5871--5886.


\bibitem{Pearce}
\newblock C. E. M. Pearce,
\newblock Quasiconvexity, fractional programming and extremal traffic congestion,
\newblock in {\em Frontiers in global optimization}, volume 74 of {\em Nonconvex Optim. Appl.}, Kluwer Acad. Publ., Boston, MA, (2004), 403--409.

\bibitem{Pel}
\newblock I. M. Pelayo,
\newblock {\em Geodesic convexity in graphs},
\newblock SpringerBriefs in Mathematics. Springer, New York, 2013.

\bibitem{Sion}
\newblock M. Sion,
\newblock On general minimax theorems,
\newblock {\em Pacific J. Math.}, \textbf{8}(1), 1958, 171--176.

\bibitem{Vel}
\newblock M. L. J. van de Vel,
\newblock {\em Theory of convex structures}, volume 50 of {\em North-Holland Mathematical Library}.
\newblock North-Holland Publishing Co., Amsterdam, 1993.



\end{thebibliography}
\end{document}